\documentclass[12pt,english]{article}
\usepackage[T1]{fontenc}
\usepackage[latin1]{inputenc}
\usepackage{babel}

\makeatletter

\providecommand{\LyX}{L\kern-.1667em\lower.25em\hbox{Y}\kern-.125emX\@}

\makeatother
\begin{document}

\title{The Casimir energy of a massive fermion field revisited}

\author{F C Santos\( ^{\dagger } \) and A C Tort\( ^{\ddagger ,\star } \)\\
 Instituto de F\'{\i}sica - Universidade Federal do Rio de Janeiro
\\
 C.P. 68528, Rio de Janeiro CEP 21945-970 Brasil}

\date{\today{}}

\maketitle
\begin{abstract}
\noindent We introduce a general, simple, and effective method of
evaluating the zero-point energy of a quantum field under the influence
of arbitrary boundary conditions imposed on the field on flat surfaces
perpendicular to a chosen spatial direction. As an example we apply
the method to the Casimir effect associated with a massive fermion
field on which MIT bag model type of boundary conditions are imposed.
\end{abstract}
\vfill
\( ^{\dagger } \) e-mail: filadelf@if.ufrj.br \\
 \( ^{\ddagger } \) e-mail: tort@if.ufrj.br \\
 \( ^{\star } \) Present address: Institut d'Estudis Espacials de
Catalunya (IEEC/CSIC) Edifici Nexus 201, Gran Capit\`{a} 2-4, 08034
Barcelona, Spain. E-mail address: visit11@ieec.fcr.es \newpage

\section{Introduction}

The macroscopically observable vacuum energy shift associated with
a quantum field is the regularised difference between the vacuum expectation
value of the corresponding hamiltonian with and without the external
conditions demanded by the particular physical situation at hand.
At the one-loop level, when the external conditions are represented
by boundary conditions this leads to the usual Casimir effect \cite{Casimir48}
-- see Ref.\cite{BMohideenM2001} for an updated review on the theoretical
and experimental aspects of this remarkable effect.

In the context of the Casimir effect, some configurations which depend
on the type of quantum field, type of spacetime manifold and its dimensionality,
specific boundary condition imposed on the quantum field on certain
surfaces, lead to relatively simple spectra, others lead to more complex
ones. The heart of the matter here is the evaluation of the spectral
sum that results from definition of the Casimir energy. This evaluation
requires regularization and renormalization, and recipes for accomplishing
this range from the relatively simple cutoff method employed by Casimir
himself \cite{Casimir48} to the powerful and elegant generalised
zeta function techniques \cite{Elisalde94}. Here, in the spirit of
the representation of a spectral sum as a contour integral \cite{relatedpapers},
we introduce a simple and effective way of evaluating the one-loop
vacuum energy under external conditions based on well-known theorems
of complex analysis, namely, the Cauchy integral formula and the Mittag-Leffler
expansion theorem in one of its simplest versions. The method we present
here is of sufficient generality so as to be successfully applied
to a variety of cases. In this letter, however, we will limit ourselves
to the example of a massive fermionic field under MIT boundary conditions
and flat surface geometry, see Ref.\cite{Johnson75} and references
therein. The massless case was first calculated by Johnson \cite{Johnson75}
and as far as the present authors are aware of there is only one evaluation
of this example of a massive fermionic Casimir energy due to Mamaev
and Trunov \cite{MamaevTrunov80}. This evaluation is only briefly
sketched by these authors, therefore it seems interesting to test
the method by providing an alternative derivation of this result.

\section{The unregularised Casimir energy and a simple sum formula}

Consider a quantum field in a \( 3+1 \) dimensional flat spacetime
under boundary conditions constraining the motion along one of the
spatial directions, say, the \( {\mathcal{OX}}_{3} \)-axis. At the
one loop-level the (unregularised) Casimir energy will be given by
\begin{equation}
\label{unregenergy}
{\mathcal{E}}=\alpha \frac{\hbar cL^{2}}{2}\int \sum _{n}\frac{d^{2}p_{\bot }}{(2\pi )^{2}}\Omega _{n},
\end{equation}
 where \( \alpha  \) is a dimensionless factor that depends on the
internal degrees of freedom of the quantum field under consideration,
\( p_{\bot }=\sqrt{p_{1}^{2}+p_{2}^{2}} \), and \begin{equation}
\Omega _{n}:=\sqrt{p_{\bot }^{2}+\frac{\lambda _{n}^{2}}{\ell ^{2}}+m^{2}},
\end{equation}
 where \( \lambda  \) is the \( n \)-th real root of the transcendental
equation determined by the boundary conditions, \( \ell  \) is a
chararacteristic length along the \( {\mathcal{OX}}_{3} \) direction,
and \( m \) is the mass of an excitation of the quantum field. A
simple integral representation of \( \sum _{n}\Omega _{n} \) can
be written if we make use of Cauchy integral formula. In fact, it
is easily seen that \begin{equation}
\label{cauchy}
\sum _{n}\Omega _{n}=-\oint _{\Gamma }\frac{dq}{2\pi }\sum _{n}\frac{2q^{2}}{q^{2}+\Omega _{n}^{2}},
\end{equation}
 where in principle \( \Gamma  \) is a Jordan curve on the \( q \)-complex
plane with \( \Im q>0 \) consisting in a semicircle of infinitely
large radius whose diameter is the entire real axis. Taking (\ref{cauchy})
into (\ref{unregenergy}) we obtain \begin{equation}
\label{unregenergy2}
{\mathcal{E}}=-\alpha \frac{\hbar cL^{2}}{2}\int \frac{d^{2}p_{\bot }}{(2\pi )^{2}}\oint _{\Gamma }\frac{dq}{2\pi }\sum _{n}\frac{2q^{2}}{q^{2}+\Omega _{n}^{2}}.
\end{equation}
 In order to proceed we must be able to perform (at least formally)
the discrete sum in (\ref{unregenergy2}).

Consider a complex function \( G(z) \) of a single complex variable
\( z \), symmetrical on the real axis such that its roots are simple,
non-zero and symmetrical with respect to the origen of the complex
plane. The assumption that the origen is not a root of \( G(z) \)
is not a restrictive one because if \( z=0 \) happens to be a root
of \( G(z) \) we can always divide \( G(z) \) by some convenient
power of \( z \) in order to eliminate zero from the set of the roots
without introducing a new singularity. Let us order and count the
roots of \( G(z) \) in such a way that \begin{equation}
\label{symmetry}
\lambda _{n}=-\lambda _{-n},\; \; \; \; \; \; n\in Z-\{0\}.
\end{equation}
 Now define the following function \begin{equation}
J(z):=\sum _{n\in Z-\{0\}}\frac{1}{z-i\lambda _{n}}.
\end{equation}
 The following properties of \( J(z) \) are self-evident: (i) \( J(z) \)
has first order poles which are determined by the roots of \( G(iz) \):
(ii) the residua are all equal to unity. Taking into account property
(\ref{symmetry}) we see that \( J(z) \) can be rewritten in the
form \begin{eqnarray}
J(z) & = & \frac{1}{2}\left( \sum _{n\in Z-\{0\}}\frac{1}{z-i\lambda _{n}}+\sum _{n\in Z-\{0\}}\frac{1}{z+i\lambda _{n}}\right) \nonumber \\
 & = & \sum _{n\in N}\frac{2z}{z^{2}+\lambda _{n}^{2}},
\end{eqnarray}
where \( Z \) denotes the set of integers and \( N \) the set of
the natural integers. Let us consider now the function \( K(z):=G(iz) \)
and state the following equality \begin{equation}
\label{J2}
J(z)=\frac{K^{\prime }(z)}{K(z)},
\end{equation}
 where the prime stands for the derivative with respect to \( z \).
In fact, the function \( K^{\prime }(z)/K(z) \) has the same simple
poles as the originally defined \( J(z) \) and the same residua,
hence we can invoke the Mittag-Leffler theorem and state that (\ref{J2})
is true. It follows then that we can write \begin{equation}
\label{sumrule}
\frac{K^{\prime }(z)}{K(z)}=\frac{d}{dz}\log K(z)=\sum _{n\in N}\frac{2z}{z^{2}+\lambda _{n}^{2}}.
\end{equation}

In order to make use of (\ref{sumrule}) we first link the dimensionless
complex variable \( z \) to the auxiliary complex momentum variable
\( q \) through the relation \begin{equation}
q^{2}+\Omega _{n}^{2}=\frac{z^{2}+\lambda _{n}^{2}}{\ell ^{2}}
\end{equation}
 hence \begin{equation}
\label{zeta}
z=\ell \sqrt{q^{2}+p_{\bot }^{2}+m^{2}},
\end{equation}
 and we can write \begin{equation}
\sum _{n}\frac{2q^{2}}{q^{2}+\Omega _{n}^{2}}=\frac{\ell q^{2}}{z}\sum _{n}\frac{2z}{z^{2}+\lambda _{n}^{2}}.
\end{equation}
 Upon changing variables (\( d/dz=(z/\ell q)d/dq \)), we obtain for
the unregularised Casimir energy the following expression \begin{equation}
{\mathcal{E}}=-\alpha \frac{\hbar cL^{2}}{2}\int \frac{d^{2}\, p_{\bot }}{(2\pi )^{2}}\oint _{\Gamma }\frac{dq}{2\pi }q\frac{d}{dq}\log K(z),
\end{equation}
 which can be integrated by parts to yield \begin{eqnarray}
{\mathcal{E}} & = & -\alpha \frac{\hbar cL^{2}}{2}\int \frac{d^{2}\, p_{\bot }}{(2\pi )^{2}}\int _{\Gamma }\frac{dq}{2\pi }\frac{d}{dq}[q\log K(z)]\nonumber \label{unregenergy3} \\
 & + & \alpha \frac{\hbar cL^{2}}{2}\int \frac{d^{2}\, p_{\bot }}{(2\pi )^{2}}\int _{\Gamma }\frac{dq}{2\pi }\log K(z).
\end{eqnarray}
 Notice that the integration is now performed on an open curve which
lies on the Riemann surface of the integrand the projection of which
on the \( q \)-complex plane is the curve \( \Gamma  \). The first
term on the R.H.S of (\ref{unregenergy3}) contributes with a phase
which cancels out with a phase coming from the second term since the
final result is real. Hence, if we keep this in mind the first term
can be put aside and the unregularised Casimir energy will be given
by \begin{equation}
\label{unregenergy4}
{\mathcal{E}}=\alpha \frac{\hbar cL^{2}}{2}\int \frac{d^{2}\, p_{\bot }}{(2\pi )^{2}}\int _{\Gamma }\frac{dq}{2\pi }\log K(z).
\end{equation}
 Equation (\ref{unregenergy4}) is our main result. It is easily seen
that it holds for arbitrary boundary conditions imposed on the field
on flat surfaces perpendicular to the \( {\mathcal{OX}}_{3} \) axis.
It also holds when the \( {\mathcal{OX}}_{3} \) direction is compactified
by the imposition of topological conditions, periodic or antiperiodic,
on the field along that direction.

\section{The massive fermion field under MIT boundary conditions}

As an example of the usefulness of (\ref{unregenergy4}) let us apply
it to a massive fermion field under MIT boundary conditions imposed
on field on the surface of an hypothetical bag consisting of two parallel
square membranes perpendicular to the \( {\mathcal{OX}_{3}} \) axis
whose side \( a \) is much larger than \( \ell  \), the distance
between them. It can be shown that for this bag model the eigenvalues
of \( p_{3} \) are determined by the roots of the function \( F(z\equiv p_{3}\ell ) \)
defined by \cite{MamaevTrunov80}\begin{equation}
F(p_{3}\ell )=\mu \sin (p_{3}\ell )+p_{3}\ell \cos (p_{3}\ell ),
\end{equation}
 where \( \mu :=m\ell  \). Therefore we can choose \( G(z) \) as
\begin{equation}
G(z)=\mu \frac{\sin z}{z}+\cos z
\end{equation}
 where we have divided \( F(z) \) by \( z \) because \( z=0 \)
is a root of \( F(z) \). We can easily prove that the roots of \( G(z) \)
are all real. Now we construct \( K(z) \) according to \begin{equation}
K(z)=G(iz)=\mu \frac{\sinh z}{z}+\cosh z.
\end{equation}
 Taking into account that for a fermionic quantum field in \( 3+1 \)
dimensions \( \alpha =-4 \), the vacuum energy is \begin{eqnarray}
{\mathcal{E}} & = & -2\hbar cL^{2}\int \frac{d^{2}p_{\bot }}{(2\pi )^{2}}\int _{\Gamma }dq\ln \left[ \frac{1}{2}\left( \frac{\mu }{z}+1\right) e^{z}\right] \nonumber \label{unregenergy5} \\
 & - & 2\hbar cL^{2}\int \frac{d^{2}p_{\bot }}{(2\pi )^{2}}\int _{\Gamma }dq\ln \left[ 1+\frac{z-\mu }{z+\mu }e^{z}\right] ,
\end{eqnarray}
 where \( z \) is defined by (\ref{zeta}). It is clear that the
first term on the R.H.S of (\ref{unregenergy5}) is divergent, however,
we can easily see also that this term is spurious in the sense that
it carries a contribution that does not depend on \( \ell  \), is
proportional to \( L^{2} \) and represents the self-energy of the
bag, plus another contribution proportional to the volume \( L^{2}\ell  \)
of the bag representing a constant energy density present even if
the bag were not there. Hence, we will subtract this term from (\ref{unregenergy5})
and write the regularised contribution as \begin{eqnarray}
{\mathcal{E}^{\makebox {(reg)}}} & = & -2\hbar cL^{2}\int \frac{d^{3}p}{(2\pi )^{3}}\ln \left[ 1+\frac{z-\mu }{z+\mu }e^{-2z}\right] \nonumber \\
 & = & -\frac{\hbar cL^{2}}{\pi ^{2}}\int _{0}^{\infty }dp\, p^{2}\ln \left[ 1+\frac{z-\mu }{z+\mu }e^{-2z}\right] .
\end{eqnarray}
 Upon changing variables again the regularised Casimir energy can
be rewritten as \begin{equation}
{\mathcal{E}^{\makebox {(reg)}}}=-\frac{\hbar cL^{2}}{\pi ^{2}\ell ^{3}}\int _{\mu }^{\infty }dz\, z\sqrt{z^{2}-\mu ^{2}}\ln \left[ 1+\frac{z-\mu }{z+\mu }e^{-2z}\right] ,
\end{equation}
 where now we have written \( z^{2}=\ell ^{2}(p^{2}+m^{2}) \) with
\( p^{2}=q^{2}+p_{\bot }^{2} \). This result is in agreement with
the result obtained in \cite{MamaevTrunov80}. For the case of massless
fermions we obtain \begin{eqnarray}
{\mathcal{E}^{\makebox {(reg)}}} & = & -\frac{\hbar cL^{2}}{\pi ^{2}\ell ^{3}}\int _{0}^{\infty }dz\, z^{3}\ln \left[ 1+e^{-2z}\right] \nonumber \label{MassFerCasEnerg} \\
 & = & -\frac{7\pi ^{2}\hbar cL^{2}}{2880\ell ^{3}},
\end{eqnarray}
 which is the result obtained in \cite{Johnson75}. Starting from
(\ref{MassFerCasEnerg}) it is a straightforward matter to obtain
the limits of the Casimir energy in the limits \( \mu \rightarrow 0 \)
and \( \mu \gg 1 \), see \cite{MamaevTrunov80}.

\section{Conclusions}

In this letter we have derived a general expression, equation (\ref{unregenergy4}),
for the evaluation of the Casimir energy of a quantum field in a flat
manifold under the influence of boundary conditions imposed on the
field on flat surfaces or topological conditions constraining the
motion along a particular spatial direction. The present authors verified
explicitly that equation (\ref{unregenergy4}) works well in several
instances, in particular, for Robin boundary conditions results agree
with those obtained by Romeo and Saharian \cite{Romeo2000}. This
method can be extended and applied to cylindrical and spherical geometries
embedded in \( d+1 \) dimensional spacetimes. Here we have chosen
the massive fermion field under MIT boundary conditions as an example
and were able to rederive its associated vacumm energy in a clear
cut way. Presently, this approach is being extended to more complex
Casimir type problems.

\section*{Acknowledgments}

One of the authors (A.C. Tort) wishes to acknowledge E. Elizalde and
the hospitality of the Institut d'Estudis Espacials de Catalunya (IEEC/CSIC)
and Universitat de Barcelona, Departament d'Estrutura i Constituents
de la Mat\`{e}ria where this work was completed, and the financial
support of CAPES, the Brazilian agency for faculty improvement, Grant
BExt 0168-01/2.


\begin{thebibliography}{1}
\bibitem{Casimir48}H.B.G. Casimir, Proc. K. Ned. Akad. Wet. \textbf{51}, (1948) 793. 
\bibitem{BMohideenM2001}M. Bordag, U. Mohideen and V.M. Mostepanenko: \textit{New Developments
in the Casimir effect}, quant-ph/0106045. 
\bibitem{Boyer68}T.H. Boyer, Phys. Rev. \textbf{174}, (1968) 1764. 
\bibitem{Elisalde94}E. Elizalde, S.D. Odintsov, A. Romeo, A.A. Bytsenko and S. Zerbini,
\textit{Zeta function regularization techniques with applications},
(World Scientific, Singapore, 1994). 
\bibitem{relatedpapers}M. Bordag, E. Elizalde and K. Kirsten, J. of Math. Phys. 37, (1996)
895: M. Bordag, E. Elizalde, B. Geyer and K. Kirsten, Comm. Math.
Phys. 179, (1996) 215; M. Bordag, E. Elizalde, K. Kirsten and S. Leseduarte,
Phys. Rev. D 56, (1997) 4896; M. Bordag, E. Elizalde and K. Kirsten,
J. of Phys. A 31, (1998) 1743.
\bibitem{Johnson75}K. Johnson, Acta Phys. Polonica \textbf{B6} (1975), 865. 
\bibitem{MamaevTrunov80}S.G. Mamaev and N.N. Trunov, Izv. Vyssh. Uchebn. Zaved., Fiz. No 7,
9 (1980). English transl. Sov. Phys. \textbf{23} (1980) 551. 
\bibitem{Romeo2000}A. Romeo and A.A. Saharian, \textit{Casimir effect for scalar fields
under Robin boundary conditions on plates}, arXiv:hep-th/0007242 v1. \end{thebibliography}
\end{document}